\documentclass[manuscript]{aastex}

\slugcomment{Submitted to  Astron J.}

\shorttitle{Small--scale systems of galaxies. II.}
\shortauthors{Gr\"utzbauch et al.}

\begin{document}

\title{Small--scale systems of galaxies. II. Properties \\ 
of the NGC~4756 group of galaxies\footnote{Based on observations obtained at
the European Southern Observatory, La Silla, Chile (Programme Nr.65.P--252)}}
\author{R.~Gr\"utzbauch} 
\affil{Institut f\"ur Astronomie, Universit\"at Wien, T\"urkenschanzstra{\ss}e 17, A-1180 Wien, Austria}
\email{gruetzbauch@astro.univie.ac.at}

\author{B.~Kelm, P.~Focardi} 
\affil{Dipartimento di Astronomia, Universit\`a di Bologna, Via Berti Pichat 6, Bologna, Italy}
\email{birgit.kelm@unibo.it, paola.focardi@unibo.it}

\author{G.~Trinchieri} 
\affil{INAF - Osservatorio Astronomico di Brera, Via Brera 28, I-20121, Milano, Italy}
\email{ginevra@brera.mi.astro.it}

\author{R.~Rampazzo} 
\affil{INAF - Osservatorio Astronomico di Padova, Vicolo dell'Osservatorio 5, I-35122, Padova, Italy}
\email{rampazzo@pd.astro.it}

\and

\author{W.W.~Zeilinger} 
\affil{Institut f\"ur Astronomie, Universit\"at Wien, T\"urkenschanzstra{\ss}e 17, A-1180 Wien, Austria}
\email{zeilinger@astro.univie.ac.at}
\begin{abstract}
This paper is part of a series that focuses on investigating 
galaxy formation and evolution in small--scale systems of galaxies
(SSSGs) in low--density environments. We present results from a study
of the NGC 4756 group which is dominated by the
elliptical galaxy NGC~5746. The characteristics of the group are investigated
through (a) the detailed investigation of the morphological,
photometric and spectroscopic properties of nine galaxies among the
dominant members of the group (b) the determination of the photometric
parameters of the faint galaxy population in an area of 34\arcmin
$\times$ 34\arcmin\ centered on NGC~4756 and (c) an 
analysis of the X-ray emission in the area based upon archival data.

The nine member galaxies are located in the core part of the NGC~4756
group (a strip diameter $\approx$ 300 kpc in diameter, H$_0$ = 70
km~s$^{-1}$~Mpc$^{-1}$) which has a very loose configuration. The
central part of the NGC~4756 group contains a significant fraction of
early--type galaxies.  Three new group members with previously unknown
systemic velocities are identified, one of which is a dE. At about
7.5\arcmin\ SW of NGC~4756 a sub--structure of the group is detected,
including IC~829, MCG~--2--33--35, MCG~--2--33--36 and
MCG~--2--33--38, which meets the Hickson criteria for being a compact
group. Most of the galaxies in this sub-structure show interaction
signatures.  We do not detect apparent fine structure and signatures of recent
interaction events in the early--type galaxy population, with the
exception of
a strong dust lane in the elliptical MCG~--2--33--38. This
galaxy displays however signatures of nuclear activity. Strong
[O~III], [N~II] and [S~II] line emission, combined with comparatively
weak, but broad H$\alpha$ emission suggest an intermediate Seyfert
type classification.

Although the area is heavily contaminated by the background cluster
Abell~1631, X--ray data suggest the presence of a hot
intergalactic medium related to the group to the X--ray emission detected. The 
present results are discussed in the context of group evolution.

\end{abstract}
\section{Introduction}

There is a growing interest in trying to identify the mechanisms
driving galaxy evolution in different kinds of environments, from
cluster cores to their outskirts and beyond, i.e. in what is commonly
called the ``field" \citep{lew02,bal03,gom03}.

Dynamical analysis and {\it n}-body simulations
\citep{bar90,bod94,moo98,dub98,Mihos99} predict that galaxies closer
than a few half-light radii, whether in a group or a rich cluster of
galaxies, are likely to merge within short timescales.  Simulations
also predict a merger remnant to display a $r^{1/4}$ profile: NGC
1132, an isolated elliptical with extended X--ray diffuse emission
\citep{mul99}, could actually represent the prototypical example of
the final evolution of all small--scale dense systems
\citep{vik99,jon03,sto03}.  Several arguments indicate however that
the chemical, dynamical and photometrical
parameters of massive ellipticals are not compatible with their
formation being the result of several major merging events diluted
along the Hubble time \citep{mez03,tho04}.  On the observational side, there
is surely good evidence of mergers at low redshift, but spectacular
systems such as the Centaurus Group do not seem to be all that common
\citep{pee03}, though they are repeatedly cited. And the
characteristics of galaxies which are observed in the act of merging
such as ULIRG systems indicate that the final product of the merging
event does not resemble a massive galaxy \citep{col01,gen01,tac02}.
 
Cluster cores certainly strongly modify the evolution of a galaxy, but
only few galaxies ($\sim$ 5\%-10\%) are cluster members.  Similarly,
it has been found that star formation is statistically enhanced in
close pairs \citep[hereafter Paper~I]{bart03,lam03,
tan03}, but again these specific environments do not host a large
fraction of the galaxy population. Most galaxies are found in groups,
which are either quite isolated or in chains or infalling into
clusters.  It is therefore important to understand whether close,
low-velocity encounters in groups are relevant mechanisms for galaxy
modification, and whether these environments are sites of the
evolution (now or in the past) for a relevant number of galaxies
\citep{car01,kel04,nik04}.

X-ray observations have greatly helped in defining the properties and
the physical reality of poor/loose groups of galaxies and compact
groups with a dominant elliptical \citep{pon96,hel00,mul00,mul03}.
Loose groups in particular are otherwise rather ``ill defined''
structures if one remains in the optical realm of observations.  One
of the more interesting results of recent years is the detection and
characterization of extended X-ray emission (an almost defining
characteristic of galaxy clusters) in groups with few accredited
members.  Even galaxy pairs \citep{hen99,tri01} are found to be
associated to diffuse X--ray emission, suggesting that systems with
different morphological characteristics and degree of compactness
represent different classes among small--scale systems of galaxies
(SSSGs hereafter).

In order to contribute to the understanding of their evolution we
started a study of SSSGs (in Paper~I), defined in 3D redshift
space. The present paper collects optical photometric and
spectro--photometric data for a sample of candidate members of the
SSSG dominated by the elliptical NGC~4756.

The goals of the present paper are the following: 1) to obtain a
redshift estimate which will provide an independent check of the
systemic velocities of the dominant NGC~4756 group members and to detect
possible faint members 2) to investigate the structure of member
galaxies through a detailed surface photometry and 3) to analyze any
possible induced activity using medium resolution spectroscopy and
diagnostic models.  The study aims also to contribute to the
detection and photometric characterization of the faint galaxie
population within the group area whose membership will be determined
by ongoing redshift surveys.

The paper is organized as follows: the literature about the NGC~4756
group is presented in \S~2. A description of the photometric and
spectroscopic observations, data reduction and analysis is given in
\S~3.  Results about photometric and spectroscopic properties of
individual bright objects, scaling relation of the candidate
early-type, from bright to dwarf, galaxy population as well as the
characteristics of the X--ray emission in the area are presented in
\S~4.

\section{The NGC~4756 group in the literature}

Before redshift surveys, NGC~4756 and its companions have been
associated to Abell~1631, a Bautz--Morgan type I cluster, within which
the morphological work of \citet{Dress80} includes 139
galaxies. Later, \citet{Dress88} compiled a catalog of radial
velocities and detected that a foreground group, well centred on the
cluster, is responsible for many of the apparently bright cluster
members. They observed that the group is much poorer than the main
cluster at cz $\approx$ 15000 km~s$^{-1}$.  The group is centred at cz
$\approx$ 4000 km~s$^{-1}$ according to \citet{Dress88} redshift
measurements.

The group LGG 306 identified by \citet{gar93} containing NGC~4756,
consists of 9 galaxies with an average velocity of 3728 km~s$^{-1}$.
According to \citet{gar93} the member candidates are IC~3799,
MCG~--2--33--17, NGC~4756, PGC~43408, PGC~43489, MCG~--2--33--35,
MCG~--2--33--36, PGC~43720 and PGC~43823. A quite different set of
galaxies associated to NGC~4756 has been given by \citet{giu00} in
their hierarchical catalog of groups. Their group 650 associates to
NGC~4756 the galaxies NGC~4714, IC~3822, IC~3827, NGC~4724, NGC~4726,
MCG~--2--33--33, NGC~4748, NGC~4782, NGC~4783, MCG~--2--33--54,
NGC~4794, NGC~4825 and MCG~--2--33--91.  At the same time NGG 4756 is
also associated to the group 664 (12 members) and to the group 675 (12
members) in the groups selected by \citet{giu00} according to another
two different percolation algorithms.  NGC 4756 is also part of the 5
member group S~188 in the UZC--SSRS2 group sample \citep{ram02} and of
the X--ray selected group SS2b189 in the \citet{mah00} sample. This
group is reported to have a velocity dispersion of 446 km~s$^{-1}$ and
a X--ray luminosity of $\log$ L$_X$=42.52 $h_{100}^{-2}$
erg~s$^{-1}$. 5 galaxies are associated to the group, rising to 8 when
all galaxies brighter than M$_{zw}$=--17 are included.

Little is known about the properties of the candidate members. Most
detailed studies concentrate on NGC~4756. RC3 \citep{rc3} classifies
the galaxy as SAB0 with a systemic velocity of 4094 $\pm$ 45
km~s$^{-1}$ from optical measurements.  \citet{Hucht94} detected
NGC~4756 in HI (V$_{sys}$ = 4280 $\pm$ 30 km~s$^{-1}$) although he
pointed out that some confusion in the signal is present due to the
nearby companions IC~829 (V$_{sys}$ = 4580 km~s$^{-1}$) at 7.5\arcmin,
MCG~--2--33--35 (V$_{sys}$ = 4123 km~s$^{-1}$) at 7.4\arcmin, and
other companions within 10\arcmin.  The inner kinematics of NGC~4756
have been measured by \cite{Pelle97}.  Within the inner 10\arcsec\ and
along the major axis (their value is P.A. = 50$^{\circ}$) the galaxy
rotation is very small (34 $\pm$ 13 km~s$^{-1}$) while the velocity
dispersion curve is nearly flat with an average value of $\langle
\sigma \rangle = $204 $\pm$ 8 km~s$^{-1}$.  X--ray emission from
NGC~4756 was first observed with the {\it Einstein Observatory} by
\citet{Fabb92}.

\section{Observation and data reduction}

The observations were carried out at the ESO 3.6m telescope equipped
with the multi--purpose instrument EFOSC2 in the period May 5 -- 7,
2000.  EFOSC2 employed a $2048 \times 2048$ pixels CCD (ESO CCD \#40).
Table~\ref{tab1} gives the coordinates of the field centers, the
exposure times and the observing conditions for the imaging part.  The
CCD was rebinned $2 \times 2$ yielding a scale of 0.314
arcsec~pixel$^{-1}$ with a field of view of $5.4 \times 5.4$
arcmin$^2$.  The NGC~4756 group has been partially covered using a
mosaic of images which allows us to obtain the photometric and
spectroscopic information of the set of galaxies labeled in
Figure~\ref{figure1}.  According to our redshift measures galaxies
labeled H1 to H6 do not belong to the group and are not considered
further in the paper. Most of the remaining galaxies visible in
Figure~\ref{figure1} have a known redshift and belong to the
background cluster.

In order to extend the mapping of galaxies in the NGC~4756 area we
analyzed a photometrically calibrated V-band image obtained with the
Wide Field Imager (WFI) at the 2.2m ESO/MPI (total FoV$\approx$34
$\times$ 34 arcmin).  The data have been originally obtained from the
WINGS group \citep{Fasano03} for the study of the background cluster
of galaxies Abell~1631. The image is a co-adding
of three non dithered frames of 460 seconds each and has a seeing of
0.7\arcsec.

\subsection{Spectroscopic observations and analysis}

A spectroscopic study of the group members was performed using both
the multi--object spectroscopy (MOS) technique and long--slit
spectroscopy.  EFOSC2 grism \#4 was used throughout the
observations. The spectral resolution was 3.36 \AA~pixel$^{-1}$.
Further details of the observations are given in Table \ref{tab2}. In
the case of long--slit spectra, the slit has been oriented along the
line connecting the nuclei of two SSSG members in order to optimize
exposure times.

The spectra were calibrated with the IRAF\footnote{IRAF is distributed
by the National Optical Astronomy Observatories, which are operated by
the Association of Universities for Research in Astronomy Inc., under
cooperative agreement with the National Science Foundation.}  software
package using standard procedures for bias subtraction and flat field
correction.  Artifacts produced by cosmic ray events were removed by
applying a filtering algorithm. Wavelength calibration was performed
on the frames by fitting a third order polynomial using as reference
the helium--argon spectrum taken before each object spectrum.  The
typical error was in the range between 0.5 to 1~{\AA}.  The spectra
were flux calibrated with the IRAF package KPNOSLIT. This procedure
included the airmass and extinction corrections using standard stars
and the atmospheric extinction coefficients published for ESO La
Silla.

Heliocentric systemic velocities have been obtained through a
cross--correlation technique using the IRAF task CROSSCOR or through
the fitting of emission or absorption lines with a single gaussian,
when the cross--correlation yielded no results because of low
signal--to--noise absorption line data.

Salient results from the spectroscopic study are given in Table~\ref{tab3}
and Table~\ref{tab4}.

\subsection{Photometric observations and analysis}

The B and R image cleaning, dark and bias subtraction, flat--fielding,
cosmic ray removal and the final calibration and image manipulation
were performed using the IRAF and ESO--MIDAS\footnote{ESO--MIDAS is
developed and maintained by the European Southern Observatory.}
software packages. Due to the large spatial extent of the group it was
necessary to make a mosaic of several images. The mosaic image of the
group, shown in Figure~\ref{figure1}, was assembled using as a
reference 7 to 10 field stars common in the different images. The
accuracy of the image alignment was usually better than 1 pixel.

Images of photometric standard stars were obtained with the same
instrumental set--up. Each of these fields contains several
calibration stars, avoiding the crowding problem typical of fields
surrounding globular cluster areas.  We used the standard photometric
transformation equations tailored to the ESO (La Silla) system of
extinction coefficients.

Ellipses were fitted by weighted least squares to the isophotes of the
NGC~4756 group member galaxies using the {\tt IRAF ISOPHOTE} package
\citep{jed87} within {\tt STSDAS}. Surface brightness, position angle
and ellipticity profiles together with the Fourier--coefficients, to
quantify the deviations of the isophotes from pure ellipses, were
derived. For the early--type members, the $b_4$ coefficient is in
particular used to analyze the boxiness/diskiness of the isophotes.
In the case of late--type disk galaxies only the surface brightness,
ellipticity and position angle profiles are used for the further
analysis. The same procedure as described in Paper~I was adopted to
analyze the surface brightness distribution in spiral galaxies.
The luminosity and geometric profiles of bright galaxies in the group
are presented in Figure~\ref{figure2}, Figure~\ref{figure3} and
Figure~\ref{figure4}. Salient data obtained from the surface
photometric study are collected in Table~\ref{tab5}. 
Results of the analysis of
the fine structure are presented in Figures~\ref{figure5} and
\ref{figure6}.

The good quality seeing of the complementary V-band Wide Field image
allowed us to detect and study the galaxy population in the field of
NGC~4576 down to faint levels. Source extraction was performed using
the {\tt SExtractor} software \citep{Bertin96}, which uses a routine
based on neural network generated weights for star-galaxy
separation. Detections were checked visually and objects with the {\it
stellaricity} parameter greater than 0.5 were deemed to be definitely
stellar and therefore discarded. All objects with a major axis
$\leq$2\arcsec\ and with a magnitude outside the range $17 \leq m_V
\leq 22$ (corresponding to $-16.8 \geq M_V \geq -11.8$) were not
processed further since they are most likely background
objects. Furtheron, all known galaxy members of the Abell~1631 cluster
were removed from the list.  Applying these selection criteria we
identify 115 galaxies as group member candidates in addition to the
objects studied with the 3.6m.

For candidate objects where neither redshift information is available
nor an unambiguous association to the NGC 4756 group is possible, we
provide their identification and the measure of their accurate
astrometric position useful for redshift followup.  The WINGS project
is indeed performing a spectroscopic study of a set of nearby clusters
up to their outskirts, including the Abell 1631 field, and will
determine therefore the group membership. Notice that \citet{zab98}
studied the galaxy population in a small number of groups. They found
a significantly higher number of faint galaxies (20-50 members down to
magnitudes as faint as M$_B \approx$ -14 +5log$_{10}$ h$_{100}$) in
X-ray detected groups, most of which are early-types, than in groups
without hot IGM. We consequently expect to find a significant
population of dwarf galaxies also in the NGC~4756 group (see X-ray
properties in \S~4.3). We further determine the accurate photometric
properties of the candidate galaxies. Several studies in the
literature start to analyze the photometric properties of the faint
galaxy population within groups \citep[see e.g.][and references
therein]{Khosro04} which could be among the driver of the group
evolution.

We used also {\tt Galfit v.2} \citep{Peng02} to obtain an accurate
modeling of the surface photometry of each galaxy which gives us the
parameters necessary to study galaxy scaling relations. In particular
we obtained the central surface brightness, $\mu_0$, the half-light
radius, r$_e$, the surface brightness at the effective radius,
$\mu_e$, and the Sersic profile shape parameter, $n$ for the bulge
component of the galaxies. 
The analysis yielded that brighter objects exhibit typical values of
$n$ in the range of $ 2 \leq n \leq 3.5$ while the faint galaxies
part of which possibly belong 
to the NGC~4756 group have values $n < 2$. The dependence of
$n$ on the absolute magnitude of  early-type galaxies has been
pointed out in several studies \citep[e.g.][]{prusi97}.
Our findings are also  in agreement with results of
similar studies e.g. such as that of \citet{Khosro04}.
The coordinates
of the bright group members (M$_B < -16.5$ mag) and the dwarf
galaxy candidates
are given in Tables~\ref{tab6} and \ref{tab7} respectively together
with the results of the surface photometry. It has to be emphasized
however, that the method of using the Sersic profile shape parameter
to identify dwarf galaxies possibly belonging to the group is prone 
to rather large uncertainties. It has to be expected therefore that
a certain fraction of the objects listed in Table~\ref{tab7} are
galaxies belonging to the background cluster. A spectroscopic
follow-up will give a definite answer.

\section{Results}

\subsection{Comments on individual galaxies from surface photometry and
spectroscopy}

The relevant data about the confirmed bright members of the NGC~4756
group are summarized in Tables \ref{tab3}, \ref{tab4} and \ref{tab5},
including three new members we found to belong to the NGC~4756 group.
Two of them are not yet catalogued and we indicated them as ``New~1"
and ``New~2" in Tables \ref{tab3} and \ref{tab5} and in
Figure~\ref{figure3} and \ref{figure6}. The third new member is Leda
83619.  In order to compare our redshift measurements we used as main
source the database produced by \citet{Dress88} in
the analysis of the background Abell cluster 1631. Seventeen objects
in \citet{Dress88} have a systemic velocity in the range of 3228 --
4943 km~s$^{-1}$. Six objects, namely MCG~--2--33--38, IC~829,
MCG~--2--33--36, PGC 43720, MCG~--2--33--35 and NGC~4756 belong to our
sample. On the average our systemic velocities compare within errors
with that of \citet{Dress88} ($\langle(our - DS88)\rangle$ = 72 $\pm$
66 km~s$^{-1}$). Only our measure of MCG~--2--33--35 is significantly
larger (192 km~s$^{-1}$) than that measured by the above authors, but
marginally consistent taking into account our measurement error.

A sub--structure composed of IC~829, MCG~--2--33--35, MCG~--2--33--36
and MCG~--2--33--38 is clearly visible SW of NGC~4756. Following
Hickson criteria \citep{hic97} this sub--structure has the
characteristics of a compact group. Indeed, the magnitudes of the
members differ by less than 3 mag, the average surface brightness of
the group is smaller than 26 mag~arcsec$^{-2}$ and there are no other
members within 3 group radii, although the dwarf galaxy New 2 is close
to this limit. At the same time, our measurements suggest that IC~829,
MCG~--2--33--35 and MCG~--2--33--38 have systemic velocities
consistent with them being part of the same group but their average
systemic velocity differs by about 1000 km~s$^{-1}$ with that of
MCG~--2--33--36 with which they seem to interact (see individual notes
of MCG~--2--33--36).

{\bf NGC~4756.}  The surface brightness profile (Figure \ref{figure2})
follows an r$^{1/4}$--law suggesting that the galaxy is a
``bona--fide'' elliptical, with an isophotal twisting of
$\approx$10$^{\circ}$ over a radial interval of 20\arcsec.  The shape
parameter $b_4$, indicates that the outer isophotes are modestly (up
to 2\%) boxy. The subtraction of the galaxy model from the original
image does not show significant residuals (Figure \ref{figure5}).  The
galaxy spectrum (Figure \ref{figure7}) does not reveal the presence of
emission lines in the wavelength region investigated (notice however,
that \citet{Cald84} reported the presence of the [O~II]$\lambda$3727
\AA\ emission feature).

{\bf IC~829.} The galaxy (Figure \ref{figure2}) is a lenticular with a
strong isophotal twisting of about 60$^{\circ}$ of the outer isophotes
probably due to the vicinity/light contamination of MCG~--2--33--36.
The color profile is typical for this class of galaxies and the
subtraction of the galaxy model to the original image does not reveal
residuals (Figure \ref{figure5}). No emission lines are detected in
the spectrum (Figure \ref{figure7}).

{\bf MCG~--2--33--38.} The surface brightness profile (Figure
\ref{figure2}) follows an r$^{1/4}$--law suggesting that the galaxy is
a ``bona--fide'' elliptical. The shape profile is irregular while the
P.A. profile does not show significant variations outside
2\arcsec. The spectrum is consistent with that of an early--type
galaxy but displays unambiguous indications of nuclear activity. Using
the data given in Table~\ref{tab4} and the diagnostic diagrams
proposed by \citet{vei87}, the galaxy nucleus is located in the region
occupied by AGNs. Strong [O~III], [N~II] and [S~II] emission combined
with comparatively weak, but broad H$\alpha$ emission (Figure
\ref{figure8}) suggests an intermediate Seyfert type classification
($\log$ [O~III]/H$\beta =0.66$, $\log$ [N~II]/H$\alpha=-0.02$, $\log$
[S~II]/H$\alpha =-0.26$, Sy type 1.9). The model subtracted image
(Figure \ref{figure5}) shows the presence of an extended dust lane
which could be the cause of the $b_3$ variation in the range of
5\arcsec $\leq r \leq$ 16\arcsec.

{\bf MCG~--2--33--36.} This spiral seems to be interacting with IC~829
and shows open outer arms/tails. After model subtraction (Figure
\ref{figure5}) a complex system of spiral arms becomes visible. The
arms starting from the galaxy nucleus are symmetric, tightly wound and
thin in the inner parts and tend to be more diffuse towards the
outskirts. The model subtraction reveals, on in the W side of the
galaxy, a third arm with a different, more open pitch angle. The
spectrum shown in Figure \ref{figure7} is typical for a Sb--type
galaxy with faint H$\alpha$ and [N~II] emission lines according to the
\citet{ken92} spectroscopic atlas of galaxies. The (B--R) color
profile with an outer average value of $\approx$1.5 is quite red for a
Sb spiral although \citet{RR96} show a few similar examples and light
contamination from IC~829 cannot be excluded.

{\bf MCG~--2--33--35.} This galaxy shows open spiral arms which depart
from a bar. This latter structure can be seen in the surface
brightness, geometric profiles (Figure \ref{figure3}) and in the
residual map (Figure \ref{figure5}). The spectrum shown in Figure
\ref{figure7} is consistent with the SBa classification of the galaxy
according to \citet{ken92}.

{\bf Leda 83169.} The galaxy is a faint and nearly edge--on late--type
spiral with a warped outer envelope and a probable inner bar structure
(see P.A. profile variation in Figure \ref{figure3}). The spectrum is
quite noisy but shows the prominent [O~II]$\lambda$3727 \AA\ feature
and possibly H$\beta$ and [O~III]$\lambda$5007 \AA\ emission lines.

{\bf New~1.}  The surface brightness profile of the galaxy shows two
components which suggest a S0 classification. The shape profile shows
moderately ($\leq$2\%) boxy isophotes in the outskirts and the
position angle profile does not reveal a significant isophotal
twisting. The color (B--R) $\approx$ 1.5 is typical of early--type
galaxies. No emission lines are visible in the very low
signal--to--noise spectrum shown in Figure \ref{figure7}. No faint
features are found after the model subtraction (Figure \ref{figure6}).

{\bf New~2.} The galaxy is a nucleated dwarf elliptical, with no other obvious
features emerging either from the surface photometry or from the faint
spectrum.

{\bf PGC 43720.} According to the surface photometry presented in Figure
\ref{figure4}, the galaxy is an early--type lenticular seen edge--on with  
small P.A. variations ($\leq$10$^{\circ}$). The residual image, after the model
subtraction, shows the disk structure mapped by the $b_4$ shape parameter.

\subsection{Group Properties}

 In our spectroscopic study, 14 galaxies in the immediate surrounding
of NGC~4756 are considered, 8 of which have redshifts accordant to the
group. Among them, we identify two new group members, one of which is
a dE, and measured the previously unknown systemic velocity of a third
member, Leda~83619. We complement our data with \citet{Dress88} and
\citet{gar93} samples in order to characterize the global group
properties.  Table~\ref{tab8} includes 11 galaxies from \citet{Dress88} with
systemic velocities in the range of 3200 -- 4950 km~s$^{-1}$ and
additionally 2 galaxies from \citet{gar93}, which are not included
either in our or in the \citet{Dress88} sample.  The member galaxies
listed by \citet{giu00} are actually too distant (on average more than
1~Mpc) from NGC~4756 to have any direct influence on NGC~4756, and we
therefore prefer not to include them in our list of group members.
The number of {\it bona-fide members} of the NGC~4756 group is then 22
galaxies.

From the above list of galaxies, the average systemic velocity of the
group is 4068 km~s$^{-1}$, with a median velocity of 4074 km~s$^{-1}$
and a velocity dispersion of 459 km~s$^{-1}$. Adopting H$_0$ = 70
km~s$^{-1}$~Mpc$^{-1}$, we derived a distance of 58 Mpc for the
NGC~4756 group. Figure~\ref{figure9} shows the distribution of the
absolute total magnitudes and of the systemic velocities of the 22
fiducial candidate members. The magnitude panel (Figure~\ref{figure9})
reveals that all galaxies belong to intermediate and low--luminosity
galaxy classes \citep{vdb98}. The (incomplete) systemic velocity
distribution (each velocity bin is 150 km~s$^{-1}$) of the group shows
that our 8 galaxies lie within a velocity range of the order of 1500
km~s$^{-1}$ a typical velocity range of loose groups once their
centers have been identified \citep{Ramel94}.

Assuming NGC~4756 as the barycenter of the group (taking also into
account its systemic velocity of 4062 $\pm$ 32 km~s$^{-1}$), we give
in Table \ref{tab8} the projected distances from this galaxy. It has
to be emphasized that the group formed by the selected 22 galaxies
(Tables \ref{tab3} and \ref{tab8}) is very sparse, in particular three
galaxies from the \citet{gar93} list are very distant in projection
from NGC~4756.

One important feature of the group is that we may isolate, at
approximately 7.5\arcmin\ SW of NGC~4756, a sub--group composed of
IC~829, MCG~--2--33--35, MCG~--2--33--36 and MCG~--2--33--38. The
sub--group fulfills all the criteria which characterize a Hickson
compact group. A long chain of galaxies (see Figure~\ref{figure1})
connects, in projection, this sub--group to NGC~4756.  Notice,
however, in the bottom panel of Figure~\ref{figure9} the high velocity
difference ($\approx$1250 km~s$^{-1}$) between the four members.

Most of the dominant members in the NGC~4756 group are disk galaxies:
lenticular (32\%) and spiral galaxies (50\%).  ``Bona--fide''
elliptical galaxies, including NGC~4756 itself and MCG~--2--33--38 (in
the sub--group), lie only in the central part of the group covered by
our study. The third early--type object, PGC 43408, classified as
peculiar elliptical in the literature, is located at 859 kpc in
projection from NGC~4756.

Presence of fine structure has been searched for in the early--type
galaxies of our sample using the scheme developed by \citet{sch92}. 
Excluding MCG~--2--33--38, however, no significant structures were
detected. The absence of fine structure may indicate that either these
galaxies have not been recently disturbed or that they have suffered weak
interactions only. The prominent dust lane in MCG~--2--33--38 is
possibly indicative of an accretion event which could also be
responsible for the refueling of the AGN engine present in this galaxy.

At variance with early--type galaxies, spirals display signatures
which could be attributed to interaction events: MCG~--2--33--36 has
multiple arms, MCG~--2--33--35 has open arms starting from the bar and
Leda 83619, seen edge--on, shows an outer warped structure and a
probable inner bar structure. According to \citet{nog86} models, open
and very regular arms may develop in a disk galaxy just after the
perigalactic passage of a perturber and can be maintained during the
very early phases of an encounter. In this context we suggest that
MCG~--2--33--36 could be distorted by the interaction with nearby
companions in the sub--group.

\subsection{Analysis of the X-ray emission in the area of the NGC~4756 group}

NGC~4756 itself deserves special attention. The galaxy lies on the
Hamabe--Kormendy relation \citep{ham87}, among bright objects, although at the
border line of the distribution of ordinary galaxies. Kinematic
observations \citep{Pelle97} suggest that the galaxy has a relatively
low central velocity dispersion value (extrapolated to 204
km~s$^{-1}$) and a very low rotation ($\leq$ 30 km~s$^{-1}$).  This
latter is a property of very bright ellipticals which, in turn, have a
higher central velocity dispersion. We attempted a measure of the central
velocity dispersion from our MOS spectroscopy using a template star
and obtained $\sigma$ = 335 $\pm$ 51 which, although biased toward a
higher value by our spectroscopic resolution, could be indicative of a
higher central velocity dispersion. Assuming the rotation velocity
given by \citet{Pelle97} and our estimate of the velocity dispersion,
the position of NGC~4756 in their Fig.~2 (upper left filled circle) would
shift towards a smaller $v_{rot}/\sigma_c$ $\approx$ 0.1, in a region
populated by galaxies associated with bright X--ray sources.  NGC~4756
is associated with an X--ray source \citep{Fabb92}, however the presence
of the background cluster Abell~1631 makes it difficult to disentangle
the contribution of the galaxy from that of the cluster.

The area of the NGC 4756 group has been observed by the ROSAT HRI and
by ASCA.  Given the complexity of the field, however neither
instrument is ideally suited for the study, because of their limited
sensitivities and for their characteristics: ASCA has a rather poor
spatial response, while the HRI, with a much better spatial
resolution, lacks the energy response that ASCA can
provide. Nevertheless, the relevant data were retrieved from the
archives\footnote{http://heasarc.gsfc.nasa.gov}.  For the HRI, the
three available observations were merged.  For ASCA only the GIS3 data
were considered. Two images representing the softer and harder energy
bands were produced from the {\it screened event} files. An adaptive
smoothing algorithm\footnote{{\it csmmoth} routine in the $CIAO$
software, see http://asc.harvard.edu} was then applied to the data.

The HRI data confirm the presence of emission localized on the
galaxies (NGC 4756 itself, MCG~--2--33--38, that hosts a Seyfert nucleus
and is located in the southern sub--group, and a bright elliptical to
the SE edge of the background cluster) but also suggests that the
large scale structure evident mostly to the N of NGC 4756 in the {\it
Einstein} data might extend also to the S (Figure~\ref{figure10}-top).
The lower angular resolution ASCA data are used to obtain images in
different energy bands. The comparison between the softer and harder
energy ones suggests that there are two seperate extended components
with slightly different spatial distributions
(Figure~\ref{figure10}-bottom): the softer band emission is
generally peaked on NGC~4756, while in the harder band the centroid
moves to the north, roughly at the position of a few galaxies in
Abell~1631.  The emission appears to extend to the southern sub-group,
however the Seyfert galaxy is likely to contribute a large fraction of
the emission observed from that region, masking the possible
contribution from hot gas local to the sub-group.

\section{Discussion and conclusions}

In an hierarchical evolutionary scenario, the group characteristics of
the diffuse X-ray emission suggest an evolutionary link between
all small-scale systems of galaxies in low density environments, from
groups to field objects. However, the link between loose groups and
isolated ellipticals is far from being demonstrated, albeit the
additional support from morphological evidence of an excess of fine
structures in isolated ellipticals \citep{Reduzzi96,Colbert01}.

We present a study of the NGC~4756 loose group, probably one of the
most difficult (observationally) but intriguing cases in the
above framework since it has in principle all the characteristics of
an ``evolving loose group''. The difficulty lies in the fact
that NGC~4756 and its companions are projected onto the
Abell~1631 cluster, although in redshift space the separation is
clear between the $cz \approx$ 15000 km~s$^{-1}$ cluster and the $cz
\approx$ 4000 km~s$^{-1}$ galaxy/group \citep{Dress88}.

Our detailed study, photometric and spectroscopic, concentrates on the
central part of the group (a strip of diameter $\approx$ 300 kpc,
H$_0$ = 70 km~s$^{-1}$~Mpc$^{-1}$) which has a very loose
configuration (see Figure 1). We extend our photometric study to the
faint galaxy population in the area in the attempt to detect and
characterize the group dwarf galaxy population.  We are aware that
there is a contamination of the background cluster but the
group/cluster membership will be refined by ongoing redshift studies.
Recent works in the literature \citep{mul99,mul00} suggest indeed the
presence of a consistent population of faint/dwarf galaxies in X--ray
bright groups which are considered a small--scale version of clusters
of galaxies. These fainter objects should mark the potential well of
the group and in a hierarchical evolutionary scenario being among the
driver of the galaxy evolution, through successive accretion
phenomena, which could trigger AGN such as that occurring in
MCG~--2--33--38 and through star formation activity re-juvenate
galaxies \citep[see e.g.][]{Weil93}.

Our work suggests that the group has at least a population of 22
galaxies within a velocity range of 1500 km~s$^{-1}$ \citep[see
e.g.][]{Ramella89} from the median of the central elliptical, NGC
4756.  At about 7.5~arcmin SW of NGC~4756 a sub--structure well
separated from NGC 4756 is detected, that  includes IC~829,
MCG~--2--33--35, MCG~--2--33--36 and MCG~--2--33--38, which meets the
Hickson criteria for being a compact group.  {\it It is noteworthy
that most of the galaxies in our sample showing interaction signatures
are concentrated in this sub-structure}.  The group then falls within
the category of loose groups which could be the birthplace of compact
groups \citep{Rood94,Ramel94}.  At the same time, the fact that the
brightest elliptical (NGC~4756) is unperturbed and quite isolated with
respect to the compact group suggests a separate history of these two
parts of the system. Recently \citet{Marcum04} analyzed a set of 8
extremely isolated E/S0 galaxies.  They suggest that at least four of
their objects are probable fossil groups.  The colors of two of them
are quite blue - (B-R)$\leq$1 - and one of them KIG~870 posseses a
possible double nucleus. None of our early-type galaxies have color
and structure characteristics similar to these objects. NGC~4756,
with its normal color and regular structure, suggests either an
unperturbed history within a pristine loose group or a very old and
``digested'' merging event. Indeed, most of the isolated objects of
\citet{Marcum04} are more similar in structure and color to NGC~4756.

At the present time only inadequate X-ray data are available and
consequently cannot allow a proper and conclusive analysis of the
X--ray properties of the NGC~4756 group.  The available X--ray data
would however point to the presence of two distinct systems at
different temperatures, corresponding to the foreground group (softer)
and the background cluster (harder).  
None of the poor groups observed in X-rays,
like those catalogued by Mulchaey et al. (2003) has optical
characteristics like those of the NGC~4756 group, 
since they are mainly either loose configurations or
compact groups from the Hickson catalogue or alike. Our data could
depict the NGC~4756 group as an ``evolving group'', consistent with
models of \citet{Diaf94,Diaf95} who proposed that compact
configurations are continually replaced after a merging episode has
taken place and produced elliptical-like members. Alternatively, the
lack of bright galaxies, in particular ellipticals, could indicate
a fundamentally different type of group compared to those having a
luminous elliptical among its members.

Combining X-ray observations with currently operating satellites with deep
optical spectroscopy of candidates dwarfs galaxies identified in this paper,
aimed at establishing their group membership, may prove to be of crucial
importance to study the potential field associated to the system, for a more
complete understanding of the properties of the NGC~4756 group of galaxies.

\acknowledgments

The authors Habib G. Khosroshahi for making available his data tables
for 16 groups.  The authors thank the anonymous referee whose comments
helped to improve the scientific content of the paper.  RG and WWZ
aknowledge the support of the Austrian Science Fund (project
P14783). RR aknowledges the kind hospitality of the Institut f\"ur
Astronomie Universit\"at Wien during the paper preparation. PF
acknowledges MIUR financial support. BK acknowledges a fellowship from
the University of Bologna.  The research has made use of the NASA/IPAC
Extragalactic Database (NED) which is operated by the Jet Propulsion
Laboratory, California Institute of Technology, under contract with
National Aeronautics and Space Administration. The Digitized Sky
Surveys were produced at the Space Telescope Science Institute under
U.S. Government grant NAG W-2166.




\begin{figure}
\caption{Mosaic of the EFOSC2 R--band images of the core region of the
NGC~4756 group. North is at the top,
East to the left. Numbers indicate galaxies considered in the present study (see
identification in Table \ref{tab3}, object no.~8 was only studied
photometrically). Galaxies indicated with the prefix H are background objects 
according to the derived systemic velocities. The
redshift of most of the other galaxies visible in the frame has been
already measured by \citet{Dress88}
\label{figure1}}
\end{figure}

\begin{figure}
\caption{ B (filled circles) and R (open circles) band surface photometry of the
confirmed NGC~4756 group members. From top to bottom: surface brightness
($\mu$), ellipticity ($\epsilon$), postion angle variation (P.A.), terms of the
Fourier expansion of residuals from the interpolation of isophotes with
ellipses (a$_3$, b$_3$, a$_4$, b$_4$) and  (B-R) color profiles.
\label{figure2}}
\end{figure}

\begin{figure}
\caption{ see Fig.~\ref{figure2}
\label{figure3}}
\end{figure}

\begin{figure}
\caption{ see Fig.~\ref{figure2}
\label{figure4}}
\end{figure}

\begin{figure}
\caption{Analysis of fine structure in galaxies: original images (left and mid
panels) and model subtracted images (right panel). From top to bottom:
NGC~4756, IC~829 + MCG~--2--33--36, MCG~--2--33--38 and MCG~--2--33--35.
\label{figure5}}
\end{figure}

\begin{figure}
\caption{Analysis of fine structure in galaxies: original images (left and mid panels) 
and model subtracted images (right panel). From top to bottom:
Leda 83619, PGC 43720, New~1, New~2.
\label{figure6}}
\end{figure}

\begin{figure}
\caption{ Spectra of galaxies in the NGC~4756 group.
\label{figure7}}
\end{figure}

\begin{figure}
\caption{ Enlargement of the MCG~--2--33--38 spectrum in order to evidence emission lines.
\label{figure8}}
\end{figure}

\begin{figure}
\caption{Distribution of absolute B-band magnitudes and systemic
velocities of the 22 galaxies we consider as members of the NGC~4756
group. The position of the confirmed group members are labeled.
Filled triangles indicate the galaxies in the sub-structure with
the characteristics of a Hickson compact group.
\label{figure9}}
\end{figure}

\begin{figure}
\caption{Isointensity X--ray contours from the HRI (top) and ASCA (bottom)
smoothed images superposed on the DSS1 image.
The HRI data are from the merged event file, for a total observing time
of $\sim$~59 ks.  The 2.5$''$ pixel image in the
PHA range 1:10 (to reduce the contribution from the particle background)
is smoothed with an adaptive smoothing algorithm.
The ASCA data are from the GIS3 instrument, for a total observing time
of $\sim$~30ks.  The two images, in two separate energy bands 1-2 keV
and 2-8 keV, are also smoothed with the same algorithm (see text).
\label{figure10}}
\end{figure}


\clearpage

\begin{deluxetable}{lcclccc} \tabletypesize{\scriptsize}
\tablecaption{Observing log of imaging. \label{tab1}}
\tablewidth{0pt} \tablehead{ \colhead{(1)} & \colhead{(2)} & \colhead{(3)}
& \colhead{(4)} & \colhead{(5)} & \colhead{(6)} & \colhead{(7)}\\ }
\startdata
1 & 12$^h$52$^m$34$^s$ & --15$^{\circ}$30$'$56$''$ & 4$\times$60  & 1.20 & 2$\times$300 & 1.19 \\
2 & 12$^h$52$^m$55$^s$ & --15$^{\circ}$23$'$48$''$ & 4$\times$60  & 1.30 & 2$\times$300 & 1.20 \\
3 & 12$^h$52$^m$44$^s$ & --15$^{\circ}$27$'$29$''$ & 3$\times$180 & 1.10 & 3$\times$180 & 1.10 \\
4 & 12$^h$52$^m$19$^s$ & --15$^{\circ}$34$'$49$''$ & 4$\times$60  & 1.21 & 2$\times$300 & 1.03 \\
5 & 12$^h$53$^m$01$^s$ & --15$^{\circ}$26$'$27$''$ & 2$\times$180 & 0.95 & 2$\times$300 & 1.07 \\
\enddata
\tablenotetext{1}{field nr.}
\tablenotetext{2}{$\alpha$ (J2000.0)}
\tablenotetext{3}{$\delta$ (J2000.0)}
\tablenotetext{4}{R-band exposure time [s]}
\tablenotetext{5}{R-band seeing FWHM [arcsec]}
\tablenotetext{6}{B-band exposure time [s]}
\tablenotetext{7}{B-band seeing FWHM [arcsec]}
\end{deluxetable}

\clearpage

\begin{deluxetable}{lccccc}
\tabletypesize{\scriptsize}
\tablecaption{Spectroscopic observations. \label{tab2}}
\tablewidth{0pt}
\tablehead{ \colhead{spectrum$^1$}   & \colhead{exp.~time} 
          & \colhead{slitwidth}      & \colhead{P.A.}  
          & \colhead{object}         & \colhead{wavelength range} \\
\colhead{ } & \colhead{[s]}    & \colhead{[arcsec]}   & \colhead{[$^\circ$]} 
            & \colhead{  }    & \colhead{[{\AA}]}  \\  }
\startdata
s286\_1 (MOS) & 2$\times$1200 & 1.8$''$ & 0$^\circ$   & MCG~--2--33--35 & 3310 -- 6330 \\
              &               &         &             & MCG~--2--33--36 & 3850 -- 7225 \\
              &               &         &             & MCG~--2--33--38 & 3790 -- 7200 \\
              &               &         &             & IC~829          & 3870 -- 7280 \\
              &               &         &             & Object~H4       & 3690 -- 6950 \\
s286\_1 (LS)  & 2$\times$1200 & 1.0$''$ & 60$^\circ$  & New~2           & 4070 -- 7470 \\
s286\_2 (MOS) & 2$\times$1200 & 1.8$''$ & 0$^\circ$   & NGC~4756        & 4350 -- 7790 \\
              &               &         &             & Leda~83619      & 2190 -- 5860 \\
              &               &         &             & Object~H2       & 3600 -- 7000 \\
              &               &         &             & Object~H3       & 3200 -- 6600 \\
s286\_4 (LS)  & 2$\times$1200 & 1.0$''$ & 137$^\circ$ & Object~H5       & 4070 -- 7470 \\
              &               &         &             & Object~H6       & 4070 -- 7470 \\
s286\_5 (LS)  & 2$\times$1200 & 1.0$''$ & 134$^\circ$ & Leda~83621~(H1) & 4070 -- 7470 \\
              &               &         &             & New~1           & 4070 -- 7470 \\
\enddata
\tablenotetext{1}{MOS -- Multi Object Spectroscopy; LS -- Longslit Spectroscopy}
\end{deluxetable}

\begin{deluxetable}{lcclllc}
\tabletypesize{\scriptsize}
\tablecaption{Spectroscopic results: systemic velocities \label{tab3}}
\tablewidth{0pt}
\tablehead{
  \colhead{object} & \colhead{$\alpha$(J2000.0)} & \colhead{$\delta$(J2000.0)} & \colhead{type}
& \colhead{V$_{hel}$ [km~s$^{-1}$]}              & \colhead{D$^1$ [kpc]}       & \colhead{id. Fig.1} \\ 
}
\startdata
MCG~--2--33--36  & 12$^h$52$^m$25.5$^s$ & --15$^{\circ}$31$'$01$''$ & Sb    & 3709 $\pm$ 160 & 158     & 4 \\
IC~829           & 12$^h$52$^m$27.4$^s$ & --15$^{\circ}$31$'$07$''$ & S0    & 4976 $\pm$ 139 & 154     & 2 \\
MCG~--2--33--35  & 12$^h$52$^m$29.4$^s$ & --15$^{\circ}$29$'$58$''$ & SBa   & 4315 $\pm$ 165 & 134     & 5 \\
MCG~--2--33--38  & 12$^h$52$^m$33.0$^s$ & --15$^{\circ}$31$'$01$''$ & E2    & 4666 $\pm$ 147 & 136     & 3 \\
New~2            & 12$^h$52$^m$41.3$^s$ & --15$^{\circ}$30$'$06$''$ & dE1,N & 4719 $\pm$ 327 & 105     & 9 \\
NGC~4756         & 12$^h$52$^m$52.6$^s$ & --15$^{\circ}$24$'$47$''$ & E3    & 4062 $\pm$ 32  & \nodata & 1 \\
New~1            & 12$^h$52$^m$54.5$^s$ & --15$^{\circ}$28$'$06$''$ & S0    & 4292 $\pm$ 126 & 59      & 7 \\
Leda~83619       & 12$^h$52$^m$59.4$^s$ & --15$^{\circ}$21$'$52$''$ & SB    & 3270 $\pm$ 133 & 58      & 6 \\
PCG~43720$^2$    & 12$^h$52$^m$51.1$^s$ & --15$^{\circ}$29$'$29$''$ & S0    & 3658 $\pm$ 45  & 82      & 8 \\
\enddata
\tablenotetext{1}{projected distance from NGC~4756}
\tablenotetext{2}{redshift from Dressler \& Shectman (1988)}
\end{deluxetable}

\begin{deluxetable}{lcccccccc}
\tabletypesize{\scriptsize}
\tablecaption{Spectroscopic results: emission line fluxes and line widths \label{tab4}}
\tablewidth{0pt}
\tablehead{
\colhead{line flux$^1$} & \colhead{H$\beta$}  & \colhead{[O~II]} & \colhead{[O~III]N1} 
                        & \colhead{[O~III]N2} & \colhead{[N~II]} & \colhead{H$\alpha$} 
                        & \colhead{[N~II]} & \colhead{[S~II]} \\
\colhead{line width$^2$} & \colhead{} & \colhead{} & \colhead{} & \colhead{} & \colhead{} 
                         & \colhead{} & \colhead{} & \colhead{} \\
}

\startdata

MCG~--2--33--36 & \nodata & \nodata & \nodata & \nodata & 4.2  &   2.8   & 11.5 & \nodata \\
                & \nodata & \nodata & \nodata & \nodata & 370  & \nodata & 233  & \nodata \\

MCG~--2--33--38 & 9.7 & \nodata & 25.2 & 43.8 & 24.4 & 61.1 & 58.1     & 18.2 \\
                & 823 & \nodata & 651  & 408  & 245  & 251  & 255$^3$  & 321  \\

Leda 83619      & 2.3 & 6.1 &    0.6  & 2.2  & \nodata & \nodata & \nodata & \nodata \\
                & 96  & 116 & \nodata & 235  & \nodata & \nodata & \nodata & \nodata \\

\enddata

\tablenotetext{1}{line fluxes are given in units of 10$^{-17}$ erg/cm$^{2}$/s/\AA}
\tablenotetext{2}{line width are in km~s$^{-1}$}
\tablenotetext{3}{width of the broad component (see Figure \ref{figure8}) could not be measured}
\end{deluxetable}

\begin{deluxetable}{lrrcccccccl}
\tabletypesize{\scriptsize}
\tablecaption{Salient properties from the surface photometry of the SSSG members
              \label{tab5}}
\tablewidth{0pt}
\tablehead{
  \colhead{member ident.}    & \colhead{B$_T$}      & \colhead{M$_B$}
& \colhead{R$_T$}            & \colhead{M$_R$}      & \colhead{r$_e$(B)}   & \colhead{r$_e$(R)}
& \colhead{$\mu_e$(B)}       & \colhead{$\mu_e$(R)} & \colhead{$\mu_0$(B)} & \colhead{$\mu_0$(R)}\\
  \colhead{(1)}     & \colhead{(2)} & \colhead{(3)}
& \colhead{(4)}     & \colhead{(5)}
& \colhead{(6)}     & \colhead{(7)} & \colhead{(8)} & \colhead{(9)} & \colhead{(10)} & \colhead{(11)}\\
}
\startdata

MCG~--2--33--36 & 14.79 & --19.03 & 13.33 & --20.49 & 8.96  & 7.67 & 21.73 & 20.07 & 19.6 & 17.7 \\
IC~829          & 14.26 & --19.56 & 12.66 & --21.16 & 5.09  & 4.95 & 20.84 & 19.22 & 18.1 & 16.2 \\
MCG~--2--33--35 & 15.99 & --17.82 & 14.34 & --19.48 & 6.13  & 6.16 & 22.65 & 21.07 & 19.9 & 18.1 \\
MCG~--2--33--38 & 14.80 & --19.02 & 13.15 & --20.66 & 8.29  & 7.89 & 22.28 & 20.57 & 19.5 & 17.6 \\
New 2           & 17.66 & --16.16 & 16.04 & --17.78 & 9.31  & 9.82 & 25.01 & 24.04 & 23.3 & 21.9 \\
PCG 43720       & 15.84 & --17.98 & 14.40 & --19.42 & 6.09  & 5.57 & 21.83 & 20.22 & 20.1 & 18.6 \\
NGC~4756        & 13.45 & --20.37 & 11.80 & --22.02 & 12.47 &11.49 & 21.53 & 19.76 & 18.8 & 17.0 \\
New 1           & 16.57 & --17.25 & 14.63 & --19.19 & 3.74  & 3.21 & 23.16 & 21.31 & 21.0 & 19.4 \\
Leda 83619      & 16.55 & --17.27 & 15.48 & --18.34 & 10.79 & 8.03 & 23.53 & 21.84 & 21.8 & 20.5 \\
\enddata
\tablenotetext{1}{SSSG and group member galaxy identification}
\tablenotetext{2}{apparent total magnitude $m_B$}
\tablenotetext{3}{absolute total magnitude $M_B$}
\tablenotetext{4}{apparent total magnitude $m_R$}
\tablenotetext{5}{absolute total magnitude $M_R$}
\tablenotetext{6}{effective radius r$_e$(B)[$''$]}
\tablenotetext{7}{effective radius r$_e$(R)[$''$]}
\tablenotetext{8}{surface brightness $\mu_e$(B) measured at the effective radius}
\tablenotetext{9}{surface brightness $\mu_e$(R)}
\tablenotetext{10}{central surface brightness $\mu_0$(B)}
\tablenotetext{11}{central surface brightness $\mu_0$(R)}
\end{deluxetable}

\begin{deluxetable}{rcccccccc}
\tabletypesize{\scriptsize}
\tablecaption{Bright group member candidates \label{tab6}}
\tablewidth{0pt}
\tablehead{
  \colhead{object$^1$}   & \colhead{$\alpha$(J2000.0)} & \colhead{$\delta$(J2000.0)}
& \colhead{class$^2$}    & \colhead{m$_V$}             & \colhead{$\mu_0$(V)} 
& \colhead{n$^3$}        & \colhead{R$_e^4$}           & \colhead{$\mu_e$(V)} \\ 
}
\startdata

      2214  &   12 51 42.3 & -15 38 53.7 & 0.02 & 16.59 & 20.16 & 1.5082           &  6.9 & 23.08 \\
      7497  &   12 53 21.7 & -15 20 48.3 & 0.03 & 18.16 & 18.89 & 0.7241           &  0.9 & 20.11 \\
      7571  &   12 53 13.0 & -15 30 12.8 & 0.03 & 16.65 & 18.51 & 1.5299           &  3.0 & 21.48 \\
      7935  &   12 53 04.6 & -15 28 48.5 & 0.03 & 17.01 & 15.62 & 2.7670           &  2.0 & 21.28 \\
      8139  &   12 52 16.3 & -15 28 43.1 & 0.02 & 22.09 & 22.37 & 0.5291           &  0.7 & 23.16 \\
      8516  &   12 53 35.6 & -15 27 49.5 & 0.03 & 18.50 & 19.35 & 1.3973           &  1.4 & 22.03 \\
      9028  &   12 52 50.8 & -15 26 42.4 & 0.03 & 17.32 & 18.43 & 1.7000           &  2.6 & 21.77 \\
     10127  &   12 51 39.2 & -15 22 26.4 & 0.00 & 20.47 & 21.32 & 0.5693           &  0.8 & 22.21 \\
     11677  &   12 52 58.2 & -15 19 50.6 & 0.03 & 18.57 & 20.18 & 0.8531           &  1.4 & 21.68 \\
     12748  &   12 53 35.6 & -15 16 14.4 & 0.03 & 14.98 & 17.12 & 2.5809           &  8.2 & 22.37 \\
     13187  &   12 51 33.6 & -15 16 20.1 & 0.03 & 17.68 & 19.44 & 1.7128           &  2.6 & 22.80 \\
     14509  &   12 52 36.7 & -15 14 04.9 & 0.03 & 16.90 & 22.19 & 0.5015           &  6.9 & 22.93 \\
     14561  &   12 52 52.4 & -15 14 32.0 & 0.03 & 16.81 & 18.63 & 1.9817           &  3.8 & 22.58 \\
     16153  &   12 52 25.3 & -15 12 22.5 & 0.03 & 18.49 & 20.96 & 8.4712           &  2.0 & 22.02 \\
     20587  &   12 53 44.8 & -15 07 42.0 & 0.03 & 17.88 & 18.61 & 3.5422           &  0.8 & 20.06 \\

\enddata
\tablenotetext{1}{SExtractor object identification}
\tablenotetext{2}{SExtractor star-galaxy separator}
\tablenotetext{3}{Sersic index n}
\tablenotetext{4}{effective radius [arcsec]}

\end{deluxetable}
\begin{deluxetable}{rcccccccc}
\tabletypesize{\scriptsize}
\tablecaption{Faint  group member candidates \label{tab7}}
\tablewidth{0pt}
\tablehead{
  \colhead{object$^1$}   & \colhead{$\alpha$(J2000.0)} & \colhead{$\delta$(J2000.0)}
& \colhead{class$^2$}    & \colhead{m$_V$}             & \colhead{$\mu_0$(V)} 
& \colhead{n$^3$}        & \colhead{R$_e^4$}           & \colhead{$\mu_e$(V)} \\ 
}
\startdata

  2132 & 12 53 48.4 & -15 38 52.6 & 0.03 &  21.38   & 21.75 & 0.4654          &  0.6 & 22.41 \\
  2604 & 12 52 25.6 & -15 38 28.5 & 0.00 &  17.57   & 20.80 & 2.6552          & 11.2 & 26.22 \\
  3379 & 12 52 19.9 & -15 36 54.9 & 0.00 &  18.82   & 21.34 & 0.9169          &  2.5 & 22.98 \\
  3529 & 12 53 00.6 & -15 37 26.1 & 0.03 &  20.42   & 20.70 & 0.5403          &  0.6 & 21.52 \\
  3981 & 12 53 41.5 & -15 36 32.4 & 0.00 &  19.79   & 23.04 & 1.1584          &  4.0 & 25.20 \\
  4115 & 12 52 36.6 & -15 36 27.5 & 0.03 &  18.54   & 18.95 & 0.7272          &  0.7 & 20.18 \\
  4129 & 12 53 21.1 & -15 36 40.7 & 0.00 &  20.60   & 20.86 & 0.5204          &  0.6 & 21.64 \\
  4542 & 12 52 32.6 & -15 36 08.5 & 0.03 &  18.79   & 19.70 & 0.8894          &  1.0 & 21.28 \\   
  5010 & 12 52 34.3 & -15 35 29.9 & 0.00 &  19.20   & 22.41 & 0.7644          &  2.4 & 23.72 \\
  5061 & 12 52 33.1 & -15 34 47.4 & 0.03 &  19.15   & 19.49 & 0.6913          &  0.6 & 20.64 \\
  5258 & 12 53 43.7 & -15 34 35.7 & 0.00 &  21.16   & 22.54 & 0.6160          &  1.2 & 23.52 \\
  5321 & 12 53 14.3 & -15 34 53.2 & 0.03 &  18.56   & 20.29 & 0.9178          &  1.8 & 21.93 \\
  5331 & 12 53 39.0 & -15 35 06.2 & 0.00 &  22.08   & 21.98 & 0.5841          &  0.6 & 22.89 \\
  5368 & 12 52 32.4 & -15 34 58.8 & 0.00 &  19.36   & 21.77 & 0.9736          &  3.7 & 23.53 \\
  5404 & 12 53 46.5 & -15 34 40.7 & 0.03 &  19.76   & 20.51 & 1.7669          &  2.5 & 23.99 \\
  5405 & 12 53 42.1 & -15 34 35.4 & 0.03 &  17.26   & 21.13 & 0.5741          &  6.3 & 22.02 \\
  5423 & 12 52 20.5 & -15 35 04.4 & 0.00 &  18.35   & 22.42 & 1.7780          & 10.2 & 25.93 \\
  5454 & 12 52 47.1 & -15 34 18.0 & 0.03 &  20.05   & 20.52 & 0.5444          &  0.6 & 21.36 \\
  5481 & 12 52 52.9 & -15 34 18.6 & 0.03 &  19.62   & 21.05 & 0.6614          &  1.3 & 22.13 \\
  5534 & 12 52 37.7 & -15 34 11.7 & 0.00 &  21.75   & 22.54 & 0.8149          &  0.9 & 23.96 \\
  5868 & 12 52 52.9 & -15 33 35.0 & 0.03 &  19.09   & 19.82 & 0.5843          &  0.8 & 20.74 \\
  5953 & 12 52 27.0 & -15 39 30.4 & 0.03 &  19.09   & 20.40 & 0.6508          &  1.3 & 21.47 \\
  6349 & 12 52 03.5 & -15 32 51.9 & 0.00 &  23.17   & 23.26 & 0.3922          &  0.6 & 23.76 \\
  6575 & 12 53 39.1 & -15 32 57.4 & 0.00 &  19.66   & 23.02 & 0.6117          &  3.4 & 24.00 \\
  6691 & 12 53 30.3 & -15 32 14.0 & 0.03 &  18.96   & 23.33 & 0.2876          &  3.1 & 23.60 \\
  6716 & 12 52 08.9 & -15 32 17.1 & 0.03 &  18.23   & 21.47 & 0.6287          &  4.4 & 22.48 \\
  6841 & 12 52 48.9 & -15 32 20.5 & 0.00 &  19.08   & 19.46 & 2.0000          &  2.1 & 23.46 \\
  7132 & 12 52 34.1 & -15 31 41.0 & 0.03 &  20.52   & 21.39 & 0.6315          &  0.9 & 22.41 \\
  7371 & 12 52 41.0 & -15 31 12.5 & 0.00 &  20.67   & 22.50 & 0.6281          &  1.3 & 23.52 \\
  7599 & 12 53 22.0 & -15 30 07.8 & 0.03 &  19.03   & 19.43 & 0.6388          &  0.6 & 20.46 \\
  7683 & 12 52 10.7 & -15 30 12.8 & 0.03 &  18.63   & 22.15 & 1.3442          &  5.8 & 24.72 \\
  7754 & 12 52 17.1 & -15 29 54.3 & 0.01 &  18.50   & 21.10 & 1.6355          &  4.1 & 24.30 \\
  8024 & 12 52 57.4 & -15 29 21.9 & 0.03 &  19.53   & 20.14 & 0.5655          &  0.7 & 21.01 \\
  8173 & 12 51 49.4 & -15 29 18.5 & 0.03 &  18.70   & 21.12 & 0.7979          &  2.7 & 22.50 \\
  8200 & 12 53 29.2 & -15 28 50.0 & 0.03 &  19.58   & 21.08 & 0.6477          &  1.4 & 22.14 \\
  8439 & 12 53 12.2 & -15 28 06.4 & 0.03 &  18.83   & 19.18 & 0.6362          &  0.7 & 20.21 \\
  8462 & 12 52 09.4 & -15 28 15.4 & 0.00 &  18.66   & 23.11 & 0.2835          &  3.6 & 23.37 \\
  8776 & 12 51 46.4 & -15 27 51.0 & 0.00 &  19.83   & 21.91 & 0.9581          &  2.6 & 23.64 \\
  8808 & 12 53 14.1 & -15 27 29.3 & 0.00 &  20.91   & 22.25 & 0.8509          &  1.1 & 23.74 \\
  8823 & 12 52 47.2 & -15 27 29.4 & 0.00 &  20.50   & 21.91 & 0.6368          &  1.4 & 22.95 \\
  8881 & 12 52 58.2 & -15 27 28.4 & 0.00 &  19.07   & 22.30 & 0.8117          &  2.9 & 23.72 \\
  9065 & 12 52 36.8 & -15 26 49.1 & 0.03 &  19.72   & 20.96 & 0.6316          &  1.4 & 21.98 \\
  9099 & 12 52 22.5 & -15 26 46.0 & 0.03 &  20.75   & 21.84 & 0.7034          &  1.0 & 23.02 \\
  9242 & 12 53 47.7 & -15 26 33.4 & 0.02 &  21.34   & 19.85 & 2.4002          &  1.3 & 24.71 \\
  9292 & 12 52 17.6 & -15 26 22.4 & 0.03 &  19.17   & 21.55 & 0.6590          &  3.1 & 22.63 \\
  9519 & 12 52 31.4 & -15 26 00.9 & 0.03 &  19.42   & 20.86 & 0.4737          &  1.4 & 21.54 \\
  9681 & 12 53 32.9 & -15 25 26.4 & 0.00 &  19.66   & 23.47 & 0.2578          &  4.5 & 23.68 \\
  9800 & 12 53 29.4 & -15 24 59.6 & 0.03 &  20.60   & 21.24 & 0.4117          &  0.6 & 21.79 \\
  9807 & 12 52 50.0 & -15 24 44.2 & 0.03 &  17.55   & 20.25 & 1.0858          &  3.0 & 22.26 \\
  9964 & 12 53 44.3 & -15 24 58.1 & 0.03 &  20.98   & 21.29 & 0.5426          &  0.6 & 22.11 \\
  9969 & 12 52 02.5 & -15 24 35.4 & 0.03 &  19.42   & 21.54 & 1.0247          &  2.8 & 23.41 \\
 10034 & 12 53 51.9 & -15 24 04.2 & 0.03 &  19.16   & 20.59 & 0.5193          &  1.0 & 21.36 \\
 10917 & 12 52 40.5 & -15 22 39.9 & 0.00 &  19.13   & 21.87 & 1.1642          &  3.7 & 24.05 \\
 10923 & 12 52 45.6 & -15 22 35.3 & 0.03 &  19.49   & 21.74 & 0.3650          &  2.1 & 22.18 \\
 10997 & 12 52 14.4 & -15 22 28.8 & 0.00 &  20.27   & 22.39 & 0.6740          &  2.1 & 23.50 \\
 11462 & 12 52 32.1 & -15 21 15.7 & 0.03 &  18.98   & 21.32 & 0.8139          &  3.2 & 22.74 \\
 11578 & 12 52 57.5 & -15 20 56.1 & 0.03 &  19.64   & 22.14 & 0.4016          &  2.4 & 22.66 \\
 11743 & 12 52 49.9 & -15 20 54.1 & 0.00 &  19.55   & 21.90 & 0.9796          &  3.4 & 23.67 \\
 11903 & 12 53 29.2 & -15 20 15.2 & 0.03 &  18.76   & 22.25 & 0.6215          &  3.3 & 23.25 \\
 12190 & 12 53 34.4 & -15 19 19.6 & 0.00 &  19.05   & 23.01 & 0.2033          &  3.4 & 23.10 \\
 12292 & 12 52 32.5 & -15 19 13.7 & 0.03 &  18.36   & 19.30 & 0.6812          &  1.0 & 20.43 \\
 12330 & 12 52 31.2 & -15 19 10.0 & 0.02 &  20.68   & 21.56 & 1.1079          &  1.4 & 23.62 \\
 12382 & 12 53 12.0 & -15 18 10.2 & 0.03 &  20.16   & 20.55 & 0.5652          &  0.6 & 21.43 \\
 12532 & 12 53 08.9 & -15 17 48.8 & 0.03 &  19.49   & 20.95 & 1.1392          &  2.1 & 23.07 \\
 12543 & 12 53 35.5 & -15 18 50.7 & 0.00 &  19.73   & 22.22 & 0.8198          &  2.4 & 23.65 \\
 12599 & 12 52 49.1 & -15 18 31.0 & 0.03 &  21.05   & 22.83 & 0.4240          &  1.3 & 23.40 \\
 12692 & 12 52 06.0 & -15 18 12.1 & 0.03 &  18.37   & 21.92 & 0.3236          &  4.1 & 22.28 \\
 12727 & 12 53 45.5 & -15 18 15.3 & 0.03 &  19.35   & 19.49 & 0.5458          &  0.6 & 20.33 \\
 12866 & 12 53 48.9 & -15 17 55.8 & 0.03 &  20.74   & 21.34 & 0.4941          &  0.7 & 22.06 \\
 12990 & 12 53 49.3 & -15 17 24.3 & 0.03 &  20.70   & 22.23 & 0.3587          &  1.1 & 22.66 \\
 13071 & 12 51 34.3 & -15 17 25.1 & 0.03 &  19.05   & 21.43 & 0.8184          &  2.1 & 22.86 \\
 13338 & 12 52 02.1 & -15 16 35.3 & 0.03 &  19.90   & 20.48 & 0.5511          &  0.7 & 21.33 \\
 13344 & 12 51 51.1 & -15 16 50.7 & 0.03 &  19.07   & 21.38 & 0.5633          &  2.7 & 22.25 \\
 13492 & 12 51 56.4 & -15 15 54.1 & 0.03 &  18.10   & 20.84 & 0.7305          &  2.6 & 22.07 \\
 13517 & 12 52 54.0 & -15 16 17.9 & 0.03 &  19.95   & 21.18 & 0.5378          &  1.1 & 21.99 \\
 13696 & 12 53 12.7 & -15 15 42.6 & 0.03 &  18.98   & 19.53 & 0.7121          &  0.8 & 20.72 \\
 14031 & 12 52 50.4 & -15 15 26.5 & 0.03 &  20.66   & 21.37 & 0.4299          &  0.7 & 21.95 \\
 14188 & 12 52 31.6 & -15 15 05.5 & 0.03 &  20.90   & 21.60 & 0.7105          &  1.1 & 22.79 \\
 14406 & 12 52 05.8 & -15 14 57.7 & 0.00 &  19.79   & 19.63 & 2.0000          &  2.1 & 23.62 \\
 14514 & 12 52 36.7 & -15 14 36.2 & 0.00 &  20.51   & 23.08 & 0.3981          &  3.4 & 23.60 \\
 14691 & 12 52 27.3 & -15 14 22.6 & 0.03 &  19.49   & 19.84 & 0.6512          &  0.6 & 20.91 \\
 14991 & 12 52 16.9 & -15 13 55.4 & 0.00 &  18.86   & 19.54 & 2.0000          &  2.1 & 23.54 \\
 15140 & 12 52 34.4 & -15 13 46.9 & 0.00 &  19.08   & 22.27 & 0.7477          &  3.0 & 23.54 \\
 15170 & 12 52 29.3 & -15 13 28.5 & 0.02 &  18.89   & 21.71 & 1.4388          &  3.9 & 24.49 \\
 15439 & 12 52 00.3 & -15 12 58.8 & 0.01 &  19.39   & 21.92 & 0.8632          &  3.5 & 23.44 \\
 15653 & 12 52 26.5 & -15 12 32.5 & 0.03 &  19.74   & 21.43 & 0.5644          &  2.2 & 22.30 \\
 15767 & 12 52 33.7 & -15 12 45.6 & 0.03 &  19.50   & 21.31 & 0.7286          &  1.9 & 22.54 \\
 16572 & 12 52 11.3 & -15 11 46.2 & 0.03 &  18.21   & 20.46 & 1.8071          &  3.5 & 24.03 \\
 16634 & 12 51 53.7 & -15 11 26.8 & 0.03 &  17.56   & 20.09 & 0.7289          &  3.2 & 21.32 \\
 17049 & 12 52 23.6 & -15 11 08.5 & 0.03 &  18.78   & 21.33 & 0.4730          &  3.0 & 22.01 \\
 17072 & 12 52 49.5 & -15 11 03.2 & 0.00 &  17.27   & 21.97 & 1.4425          &  8.2 & 24.74 \\
 17082 & 12 52 24.8 & -15 10 51.5 & 0.03 &  19.55   & 19.92 & 0.6656          &  0.7 & 21.01 \\
 17695 & 12 52 30.6 & -15 10 21.3 & 0.00 &  19.06   & 21.66 & 1.0611          &  3.6 & 23.61 \\
 18044 & 12 52 16.8 & -15 10 15.9 & 0.00 &  19.54   & 21.37 & 0.7604          &  1.5 & 22.67 \\
 18609 & 12 52 35.8 & -15 09 43.8 & 0.00 &  19.43   & 18.87 & 2.0000          &  1.8 & 22.86 \\
 18624 & 12 52 01.9 & -15 09 45.9 & 0.00 &  21.70   & 22.57 & 1.1557          &  1.5 & 24.73 \\
 19459 & 12 51 40.2 & -15 08 46.9 & 0.00 &  20.54   & 21.80 & 0.7618          &  1.1 & 23.10 \\
 19612 & 12 52 54.6 & -15 09 28.0 & 0.00 &  18.19   & 21.87 & 1.4264          &  7.8 & 24.61 \\
 19679 & 12 53 21.9 & -15 08 06.7 & 0.03 &  19.10   & 19.61 & 0.9773          &  0.8 & 21.38 \\
 20203 & 12 51 41.1 & -15 07 25.5 & 0.03 &  18.31   & 19.69 & 1.1262          &  1.9 & 21.78 \\

\enddata
\tablenotetext{1}{SExtractor object identification}
\tablenotetext{2}{SExtractor star-galaxy separator}
\tablenotetext{3}{Sersic index n}
\tablenotetext{4}{effective radius [arcsec]}
\end{deluxetable}

\begin{deluxetable}{lccllllc}
\tabletypesize{\scriptsize}
\tablecaption{Additional group member candidates not included in our sample \label{tab8}}
\tablewidth{0pt}
\tablehead{
  \colhead{object}   & \colhead{$\alpha$(J2000.0)} & \colhead{$\delta$(J2000.0)}
& \colhead{type}     & \colhead{m$_B$ [mag]}       & \colhead{cz [km~s$^{-1}$]} 
& \colhead{D$^1$ [kpc] } & \colhead{Notes$^2$} \\ }
\startdata
PCG~43966       & 12$^h$54$^m$49.2$^s$ & --16$^{\circ}$03$'$06$''$ & SB(s)m          & 15    & 3400 $\pm$ 45 & 812 & DS-4      \\
PCG~43808       & 12$^h$53$^m$33.4$^s$ & --15$^{\circ}$52$'$44$''$ & S               & 15    & 4386 $\pm$ 45 & 478 & DS-12     \\
PCG~43505       & 12$^h$51$^m$00.4$^s$ & --15$^{\circ}$40$'$50$''$ & S0              & 16    & 4255 $\pm$ 45 & 547 & DS-26     \\
PCG~43711       & 12$^h$52$^m$48.9$^s$ & --15$^{\circ}$35$'$21$''$ & S0              & 15    & 4432 $\pm$ 45 & 179 & DS-28     \\
MCG~--2--33--33 & 12$^h$52$^m$05.4$^s$ & --15$^{\circ}$27$'$31$''$ & Sb              & 15    & 4424 $\pm$ 45 & 204 & DS-54     \\
PCG~43903       & 12$^h$54$^m$22.8$^s$ & --15$^{\circ}$23$'$59$''$ & SBa             & 16    & 4402 $\pm$ 45 & 380 & DS-55     \\
PCG~43823       & 12$^h$53$^m$42.3$^s$ & --15$^{\circ}$16$'$56$''$ & S0              & 14    & 4085 $\pm$ 45 & 248 & DS-75; G  \\
PCG~44021       & 12$^h$55$^m$19.3$^s$ & --14$^{\circ}$56$'$59$''$ & Sd              & 15.2  & 3228 $\pm$ 45 & 777 & DS-121    \\
PCG~43489       & 12$^h$50$^m$52.3$^s$ & --14$^{\circ}$54$'$24$''$ & Sc              & 15    & 3707 $\pm$ 45 & 722 & DS-125; G \\
PCG~43913       & 12$^h$54$^m$27.5$^s$ & --14$^{\circ}$42$'$13$''$ & S0/a            & 14    & 3945 $\pm$ 45 & 823 & DS-136    \\
MCG~--2--33--17 & 12$^h$50$^m$04.4$^s$ & --14$^{\circ}$44$'$06$''$ & (R)SA(r)0+ pec? & 12.90 & 3976 $\pm$ 41 & 987 & G         \\
PCG~43408       & 12$^h$49$^m$52.9$^s$ & --15$^{\circ}$00$'$52$''$ & Ep              & 14    & 3902 $\pm$ 45 & 859 & DS-116; G \\
IC~3799         & 12$^h$48$^m$59.6$^s$ & --14$^{\circ}$23$'$57$''$ & Sd              & 14    & 3693 $\pm$ 9 & 1421 & G         \\
\enddata
\tablenotetext{1}{projected distance from NGC~4756}
\tablenotetext{2}{DS -- galaxy identification from Dressler \& Shectman (1988); G -- group members of LGG 306 according to Garcia (1993)}
\end{deluxetable}


\begin{thebibliography}{}
\bibitem[Balogh et al.(2003)]{bal03} Balogh, M.L. et al. 2003, astro-ph/0311379
\bibitem[Barnes(1990)]{bar90} Barnes, J.E. 1990, Nature, 344, 379
\bibitem[Barton et al.(2003)]{bart03} Barton, G.E., Geller, M.J., Kenyon, S.J. 2003,
        \apj ~{\bf 582}, 668
\bibitem[Bertin \& Arnouts(1996)]{Bertin96} Bertin, G., Arnouts, S. 1996, A\&AS ~{\bf 117}, 393
\bibitem[Bode et al.(1994)]{bod94} Bode, P.W., Berrington, R.C., Cohn, H.N., Lugger, P.M. 1994,
        \apj ~{\bf 433}, 479
\bibitem[Caldwell (1984)]{Cald84} Caldwell, N. 1984, \pasp ~{\bf 96}, 287
\bibitem[Carlberg et al.(2001)]{car01} Carlberg, R.G., Yee, H.K.C., Morris, S.L. et al. 2001,
        \apj ~{\bf 563}, 736
\bibitem[Colbert et al.(2001)]{Colbert01} Colbert, J.W., Mulchaey, J.S., Zabludoff, A.I. 2001, \aj~{\bf 107}, 2035.
\bibitem[Colina et al.(2001)]{col01} Colina, L. et al. 2001, \apj ~{\bf 563}, 546
\bibitem[de Vaucouleurs et al.(1991)]{rc3} de Vaucouleurs, G., de Vaucouleurs, A., Corwin, H.G. Jr., Buta, R.J., Paturel, G. Fouqu\'e P. 1991, Third Reference Catalog of Bright Galaxies (New York: Springer Verlag) 
\bibitem[Diaferio et al.(1994)]{Diaf94} Diaferio, A., Geller, M.J., Ramella, M. 1994, \aj~{\bf 107}, 868
\bibitem[Diaferio et al.(1995)]{Diaf95} Diaferio, A., Geller, M.J., Ramella, M. 1995, \aj~{\bf 109}, 2293
\bibitem[Dressler(1980)]{Dress80} Dressler, A., 1980, \apjs ~{\bf 42}, 565
\bibitem[Dressler \& Shectman(1988)]{Dress88} Dressler, A., Shectman, S.A., 1988, \aj ~{\bf 95}, 284
\bibitem[Dubinski(1998)]{dub98} Dubinski, J., 1998, \apj ~{\bf 502}, 141
\bibitem[Fabbiano, Kim \& Trinchieri(1992)]{Fabb92} Fabbiano, G., Kim, D.W., Trinchieri, G. 1992,  \apjs ~{\bf 80}, 531
\bibitem[Fasano et al.(2003)]{Fasano03} Fasano, G., Poggianti, B., Bettoni, D., Pignatelli, E., Marmo, C., Moles, M., Kj\"ajgaard, P., Varela, J., Couch, W., Dressler, A. 2003, MemSAI ~{\bf 74}, 355
\bibitem[Garcia(1993)]{gar93} Garcia, A.M. 1993, \aaps ~{\bf 100}, 47
\bibitem[Genzel et al.(2001)]{gen01} Genzel, R., Tacconi, L.J., Rigopoulou, D., Lutz, D., Tecza, M.,
        2001, \apj ~{\bf 563}, 527
\bibitem[Giuricin et al.(2000)]{giu00} Giuricin, G., Marinoni, C., Ceriani, L., Pisani, A. 2000, 
        \apj ~{\bf 543}, 178.
\bibitem[Gomez et al.(2003)]{gom03} Gomez, P.L. et al. 2003, \apj ~{\bf 584}, 210
\bibitem[Hamabe \& Kormendy(1987)]{ham87} Hamabe, M.,  Kormendy, J. 1987, 
        {\it Structure and Dynamics of elliptical Galaxies}, IAU Symp. No. 127 (Princeton), 
        ed. T. de Zeeuw, Dordrecht: Reidel, 379.
\bibitem[Helsdon \& Ponman(2000)]{hel00} Helsdon, S.F., Ponman, T.J. 2000, \mnras ~{\bf 319}, 933
\bibitem[Henricksen \& Cousineau(1999)]{hen99} Henricksen, M., Cousineau, S. 1999, \apj ~{\bf 511}, 595
\bibitem[Huchtmeier(1994)]{Hucht94} Huchtmeier, W.K. 1994, \aap ~{\bf 286}, 389
\bibitem[Hickson(1997)]{hic97} Hickson, P. 1997, \araa ~{\bf 35}, 357
\bibitem[Jedrzejewski(1987)]{jed87} Jedrzejewski, R. 1987, \mnras ~{\bf 226}, 747
\bibitem[Jones et al.(2003)]{jon03} Jones, L.R., Ponman, T.J., Horton, A., Babul, A., Ebeling, H., Burke, D.J.
        2003, \mnras ~{\bf 343}, 627
\bibitem[Kelm \& Focardi(2004)]{kel04} Kelm, B., Focardi, P. 2004, astro-ph/0411135
\bibitem[Kennicutt(1992)]{ken92} Kennicutt, R.C. 1992, \apjs ~{\bf 79}, 255
\bibitem[Khosroshahi et al.(2004)]{Khosro04} Khosroshahi, H.G., Raychaudhury S., 
Ponman, T.J., Miles T.A., Forbes, D.A. 2004, MNRAS 349, 527
\bibitem[Lambas et al.(2003)]{lam03} Lambas, D.G.,Tissera, P.B., Sol Alonso, M., Coldwell, G. 2003,
        \mnras ~{\bf 346}, 1189
\bibitem[Lewis et al.(2002)]{lew02} Lewis, I. et al. 2002, \mnras ~{\bf 334}, 673
\bibitem[Mahdavi et al.(2000)]{mah00} Mahdavi, A., B\"ohringer, H., Geller, M.J., Ramella, M.
        2000, \apj ~{\bf 534}, 114
\bibitem[Marcum et al.(2004)]{Marcum04} Marcum, P.M., Aars, C.E., Fanelli M.N. 2004, 
        \aj ~{\bf 127}, 3213
\bibitem[Meza et al.(2003)]{mez03} Meza, A., Navarro, J.F., Steinmetz, M., Eke, V.R. 2003, \apj ~{\bf 590}, 619
\bibitem[Mihos (1999)]{Mihos99} Mihos, C. 1999, Astro. and Space Science {\bf 266}, 195
\bibitem[Moore et al.(1998)]{moo98} Moore, B., Lake, G., Katz, N. 1998, \apj ~{\bf 495}, 139
\bibitem[Mulchaey \& Zabludoff(1999)]{mul99} Mulchaey, J. S., Zabludoff, A. I. 1999, \apj ~{\bf 514}, 33
\bibitem[Mulchaey(2000)]{mul00} Mulchaey, J. S. 2000, \araa ~{\bf 38}, 289
\bibitem[Mulchaey et al.(2003)]{mul03} Mulchaey, J., Davis, D.S., Mushotzky, R.F., Burstein, D. 2003, \apjs ~{\bf 145}, 39
\bibitem[Navarro(1990)]{nav90} Navarro, J. F. 1990, \mnras ~{\bf 242}, 311
\bibitem[Nikolic et al.(2004)]{nik04} Nikolic, B., Cullen, H., Alexander, P. 2004, 
        \mnras ~in press (astro-ph/047289)
\bibitem[Noguchi \& Ishibashi(1986)]{nog86} Noguchi, M., Ishibashi, S. 1986, \mnras ~{\bf 219}, 305
\bibitem[Peebles(2003)]{pee03} Peebles, J.E. 2003, astro-ph/0309269
\bibitem[Peng et al.(2002)]{Peng02} Peng, C.Y.; Ho, L.C., Impey, C.D., Rix, H.-W. 2002, \aj~{\bf 124}, 266
\bibitem[Pellegrini et al.(1997)]{Pelle97} Pellegrini, S., Held, E.V., Ciotti, L. 1997, \mnras ~{\bf 288}, 1
\bibitem[Ponman et al.(1996)]{pon96} Ponman, T. J., Bourner, P. D. J., Ebeling, H., B\"ohringer, H. 1996, 
        \mnras ~{\bf 283}, 690
\bibitem[Prugniel \& Simien(1997)]{prusi97} Prugniel, P., Simien, F. 1997, \aap ~{\bf 321}, 111
\bibitem[Ramella et al.(1989)]{Ramella89} Ramella, M. Geller, M.J.; Huchra, J.P. 1989 \apj~{\bf 344}, 57 
\bibitem[Ramella et al.(1994)]{Ramel94} Ramella, M., Diaferio, A., Geller, M.J., Huchra, J.P. 1994, 
        \aj~{\bf 107}, 1623        
\bibitem[Ramella et al.(2002)]{ram02} Ramella, M., Geller, M.J., Pisani, A., da Costa, L.N. 2002,.
        \aj ~{\bf 123}, 2976
\bibitem[Rampazzo et al.(2000)]{ram00} Rampazzo, R., D'Onofrio, M., Bonfanti, P., Longhetti, M., Reduzzi, L. 
        2000, \aplett ~{\bf 40}, 63
\bibitem[Reduzzi \& Rampazzo(1996)]{RR96} Reduzzi, L., Rampazzo, R. 1996, \aaps ~{\bf 116}, 515
\bibitem[Reduzzi et al.(1996)]{Reduzzi96} Reduzzi, L., Longhetti, M., Rampazzo, R. 1996, \mnras~{\bf 282}, 149
\bibitem[Rood \& Struble(1994)]{Rood94} Rood, H.J., Struble, M.F. 1994, PASP~{\bf 106}, 413
\bibitem[Schweizer(1992)]{sch92} Schweizer, F., 1992, 
        {\it Structure, Dynamics and Chemical Evolution of Early--type Galaxies}, 
        ESO--EIPC Workshop, Danziger et al., 651
\bibitem[Stocke et al.(2003)]{sto03} Stocke, J.T., Keeney, B.A., Lewis, A.D., Epps, H.W., Schild, R.E.
        2003, astro-ph/0311572
\bibitem[Tacconi et al.(2002)]{tac02} Tacconi, L.J., Genzel, R., Lutz, D. 2002, \apj ~{\bf 580}, 73
\bibitem[Tanvuia et al.(2003)] {tan03} Tanvuia, L., Kelm, B., Focardi, P., Rampazzo, R., Zeilinger, W.W. 
        2003, \aj ~{\bf 126}, 1245 (Paper~I)
\bibitem[Thomas et al.(2004)]{tho04} Thomas, D., Maraston, C., Bender, R., Mendes de Olivera, C.,
        2004, \mnras ~accepted (astro-ph/0410209)
\bibitem[Trinchieri \& Rampazzo(2001)]{tri01} Trinchieri, G., Rampazzo, R. 2001, \aap ~{\bf 374}, 454
\bibitem[van den Bergh(1998)]{vdb98} van den Bergh, S. 1998, {\it Galaxy morphology and classification}, 
        Cambridge University Press
\bibitem[Veilleux \& Osterbrock(1987)]{vei87} Veilleux, S., Osterbrock, D. E. 1987, \apjs ~{\bf 63}, 295
\bibitem[Vikhlinin et al.(1999)]{vik99} Vikhlinin, A. et al. 1999, \apj ~{\bf 520}, L1
\bibitem[Weil \& Hernquist(1993)]{Weil93} Weil, M.L., Hernquist, L. 1993, \apj~{\bf 405}, 142
\bibitem[Zabludoff \& Mulchaey(1998)]{zab98} Zabludoff, A., Mulchaey, J.S. 1998, \apj ~{\bf 496}, 39

\end{thebibliography}
\end{document}